\documentclass
[twocolumn,showpacs,preprintnumbers,superscriptaddress,amsmath,amssymb,prc]{revtex4}
\usepackage{graphicx}
\usepackage{dcolumn}
\usepackage{bm}

\begin{document}

\title{ New measurement of the excited states in $^{11}$B via the elastic resonance scattering of $^{10}$Be + p}

\author{ Y. D. Liu}
\affiliation{Shanghai Institute of Applied Physics, Chinese Academy of Sciences, Shanghai 201800, China}
\affiliation{University of Chinese Academy of Sciences, Beijing 100049, China}
\author{ H. W. Wang}
\affiliation{Shanghai Institute of Applied Physics, Chinese Academy of Sciences, Shanghai 201800, China}

\author{Y. G. Ma}
\thanks{Author to whom all correspondence can be addressed. Email: ygma@sinap.ac.cn}
\affiliation{Shanghai Institute of Applied Physics, Chinese Academy of Sciences, Shanghai 201800, China}
\affiliation{Shanghai Tech University, Shanghai 200031, China}

\author{ X. G. Cao}
\affiliation{Shanghai Institute of Applied Physics, Chinese Academy of Sciences, Shanghai 201800, China}
\author{ G. Q. Zhang}
\affiliation{Shanghai Institute of Applied Physics, Chinese Academy of Sciences, Shanghai 201800, China}
\author{ D. Q. Fang}
\affiliation{Shanghai Institute of Applied Physics, Chinese Academy of Sciences, Shanghai 201800, China}
\author{ M. Lyu}
\affiliation{Shanghai Institute of Applied Physics, Chinese Academy of Sciences, Shanghai 201800, China}
\affiliation{University of Chinese Academy of Sciences, Beijing 100049, China}
\author{ W. B. He}
\affiliation{Shanghai Institute of Applied Physics, Chinese Academy of Sciences, Shanghai 201800, China}
\affiliation{University of Chinese Academy of Sciences, Beijing 100049, China}
\author{ Z. T. Dai}
\affiliation{Shanghai Institute of Applied Physics, Chinese Academy of Sciences, Shanghai 201800, China}
\affiliation{University of Chinese Academy of Sciences, Beijing 100049, China}
\author{ C. Li}
\affiliation{Shanghai Institute of Applied Physics, Chinese Academy of Sciences, Shanghai 201800, China}
\author{ C. L. Zhou}
\affiliation{Shanghai Institute of Applied Physics, Chinese Academy of Sciences, Shanghai 201800, China}
\author{S. Q. Ye}
\affiliation{Shanghai Institute of Applied Physics, Chinese Academy of Sciences, Shanghai 201800, China}
\affiliation{University of Chinese Academy of Sciences, Beijing 100049, China}
\author{ C. Tao}
\affiliation{Shanghai Institute of Applied Physics, Chinese Academy of Sciences, Shanghai 201800, China}
\affiliation{University of Chinese Academy of Sciences, Beijing 100049, China}

\author{ J. Wang}
\affiliation{Shanghai Institute of Applied Physics, Chinese Academy of Sciences, Shanghai 201800, China}
\affiliation{University of Chinese Academy of Sciences, Beijing 100049, China}
\author{ S. Kumar }
\affiliation{Shanghai Institute of Applied Physics, Chinese Academy of Sciences, Shanghai 201800, China}

\author{ J. L. Han}
\affiliation{Institute of Modern Physics, Chinese Academy of Sciences, Lanzhou 730000, China}
\author{ Y. Y. Yang}
\affiliation{Institute of Modern Physics, Chinese Academy of Sciences, Lanzhou 730000, China}
\affiliation{University of Chinese Academy of Sciences, Beijing 100049, China}
\author{ P. Ma}
\affiliation{Institute of Modern Physics, Chinese Academy of Sciences, Lanzhou 730000, China}
\author{ J. B. Ma}
\affiliation{Institute of Modern Physics, Chinese Academy of Sciences, Lanzhou 730000, China}
\author{ S. L. Jin}
\affiliation{Institute of Modern Physics, Chinese Academy of Sciences, Lanzhou 730000, China}
\affiliation{University of Chinese Academy of Sciences, Beijing 100049, China}
\author{ Z. Bai}
\affiliation{Institute of Modern Physics, Chinese Academy of Sciences, Lanzhou 730000, China}
\affiliation{University of Chinese Academy of Sciences, Beijing 100049, China}
\author{ L. Jin}
\affiliation{Institute of Modern Physics, Chinese Academy of Sciences, Lanzhou 730000, China}
\author{ D. Yan}
\affiliation{Institute of Modern Physics, Chinese Academy of Sciences, Lanzhou 730000, China}
\author{ J. S. Wang}
\affiliation{Institute of Modern Physics, Chinese Academy of Sciences, Lanzhou 730000, China}

\date{ \today}

\begin{abstract}
The elastic resonance scattering protons decayed from $^{11}$B to the ground state of $^{10}$Be
were measured using the thick-target technique in inverse kinematics at the Heavy Ion
Research Facility in Lanzhou (HIRFL). The obtained excitation functions were well described
by a multichannel R-matrix procedure under the kinematics
process assumption of resonant elastic scattering. The excitation energy of the resonant states ranges from 13.0 to 17.0 MeV,
and their resonant parameters such as the resonant energy E$_{x}$, the spin-parity J$^\pi$, and the proton-decay partial width $\Gamma_p$
were determined from R-matrix fits to the data. Two of these states around
E$_{x}$ = 14.55 MeV [J$^\pi$ = (3/2$^+$, 5/2$^+$), $\Gamma_p$ = 475 $\pm$ 80 keV] and E$_{x}$ =
14.74 MeV [J$^\pi$ = 3/2$^-$, $\Gamma_p$ = 830 $\pm$ 145 keV], and a probably populated state
at E$_x$ = 16.18 MeV [J$^\pi$ =(1/2$^-$, 3/2$^-$), $\Gamma_p$ $<$ 60 keV], are respectively
assigned to the well-known states in $^{11}$B at 14.34 MeV, 15.29 MeV, and 16.43 MeV.
The isospin of these three states were previously determined to be T = 3/2, but discrepancies exist
in widths and energies due to the current counting statistics and energy resolution.
We have compared these states with previous measurements,
and the observation of the possibly populated resonance is discussed.
\end{abstract}
\maketitle

\section{Introduction}
\label{sec:Intro}
\begin{figure}[htbp]
\resizebox{8.6cm}{!}{\includegraphics{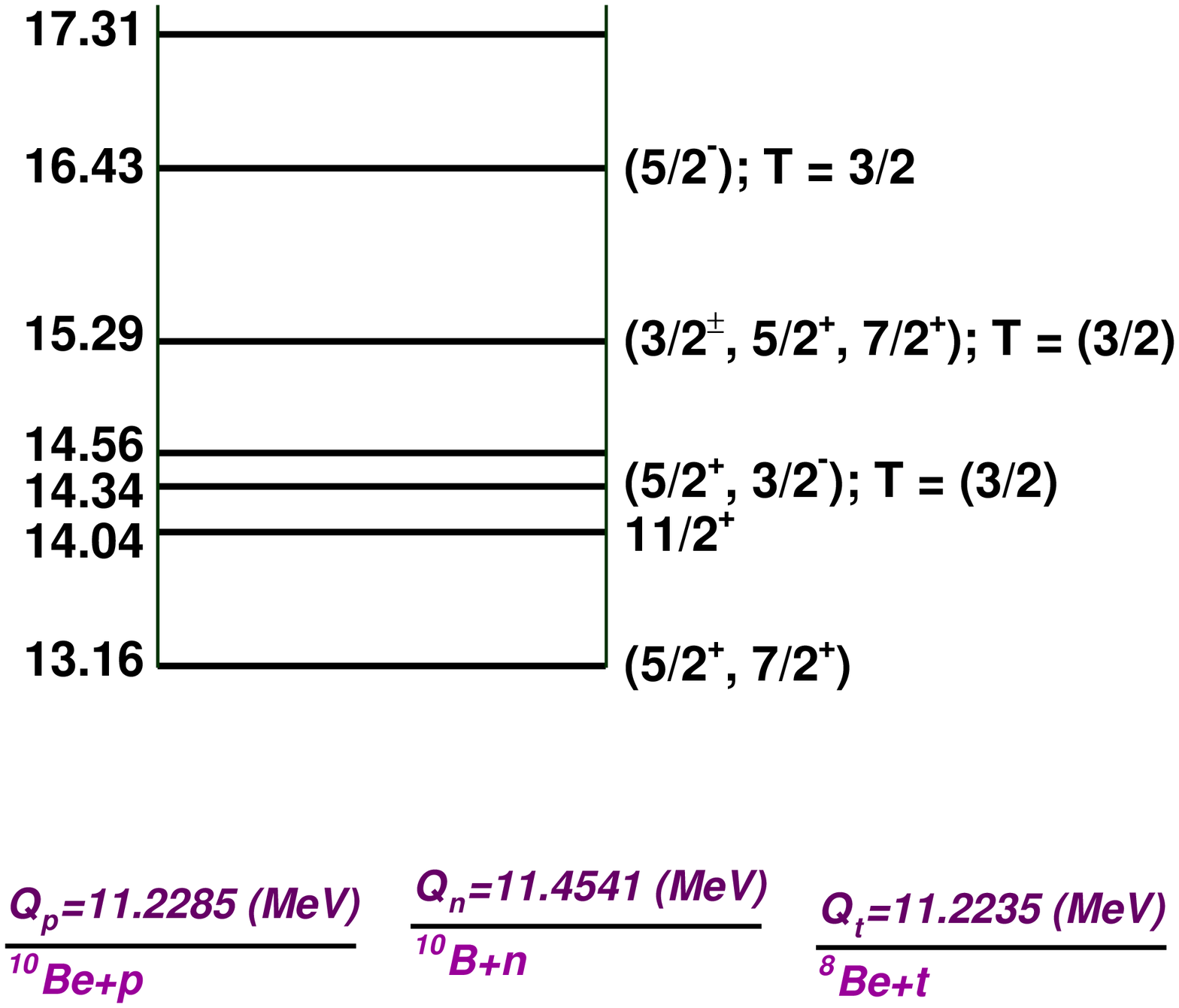}}
\vspace{-0.8cm}
\caption{(Color online) The partial energy level diagram of  $^{11}$B from Refs.~\cite{Ajzenberg1975, Ajzenberg1985, Kelly2012}, with energy levels in MeV and the possible spin-parities and isospin component (T) shown in parenthesis.}
\label{b11levelspre}
\vspace{+0.0cm}
\end{figure}

As one of the most intriguing fields in nuclear physics with available heavy ion beams, the
exploration of resonant structure especially from an alpha-clustering perspective in light nuclei has
attracted much attention in recent years \cite{Freer2007, Oertzen2006, He2014, He2014N, Tao2013, Ye2014}.
Measurements in inverse kinematics, for instance, were made to determine the exotic structure or decay mode of the compound
nuclei $^{18}$F, $^{18}$Ne, $^{11,14}$C, $^{11}$N \cite{Bailey2014, Campo2000, Freer2012, Freer2014, Markenroth2000}.
For the $^{11}$B nucleus, which was identified as an alpha-clustering compound system at several low-lying excited states,
there are many measurements using different reactions endeavored for extracting
the information of resonant structure and energy levels, while the ambiguousness for its high excitation state properties of
isospins T = 3/2 and T = 1/2 still remains. Just as shown in Figure \ref{b11levelspre}, little spectroscopic
knowledge of unbound excited states in $^{11}$B has been determined due to the difficulty of extracting the parameters of states
above the separation energies of proton (Q$_p$ = 11.2285 MeV), neutron (Q$_n$ = 11.4541 MeV) and
triton (Q$_t$ = 11.2235 MeV), where several particle channels are simultaneously open. As the analog of the 1.78 MeV state in $^{11}$Be, primary evidence for the 14.33 MeV state with $J^\pi$ = ($5/2^+$, $3/2^-$), T = 3/2 assignment comes from the $^{10}$Be(p, $\gamma$)$^{11}$B reaction \cite{Goosman1973},
which also yielded the population of 15.30 MeV, the analog state of 2.70 MeV in $^{11}$Be, but without spin-parity and T assignment. For the 15.30 MeV state, there is no definite $J^\pi$ information reported in Refs. \cite{Zwieglinski1979, Zwieglinski1982, Soic2004, Watson1971, Fletcher2003}, and its width and spin-parity determinations from the work \cite{Goosman1973, Aryaeinejad1985, Sadowski1990} are also questionable just as demonstrated by Refs. \cite{Fortune2006, Barker2007}.

In the present work, the measurement was undertaken in the hope
that the properties, i.e., the excitation energy E$_{x}$, the spin-parity J$^\pi$, and the proton-decay partial width of those unbound excited states lied over the energy region of present interest shown in Fig. \ref{b11levelspre} could be extracted and determined by employing the elastic resonance reaction $^{10}$Be + p with the thick-target technique in inverse kinematics
(TTIK) \cite{Artemov1990, Uribarri2000, Hernandez1998}. Comparing with those traditional approaches,
the TTIK method takes advantage of the fact that a continuous excitation function spectrum
can be achieved when the incident energy of beam particles on the target remains unchanged. In the following, the experiment details are described in Sec. \ref{sec:expintro}, the results and discussion are presented in Sec. \ref{sec:resultanddiss}, and finally the summary is given in Sec. \ref{sec:sum}.

\section{details of the experiment}
\label{sec:expintro}

\begin{figure}[htbp]
\resizebox{8.6cm}{!}{\includegraphics{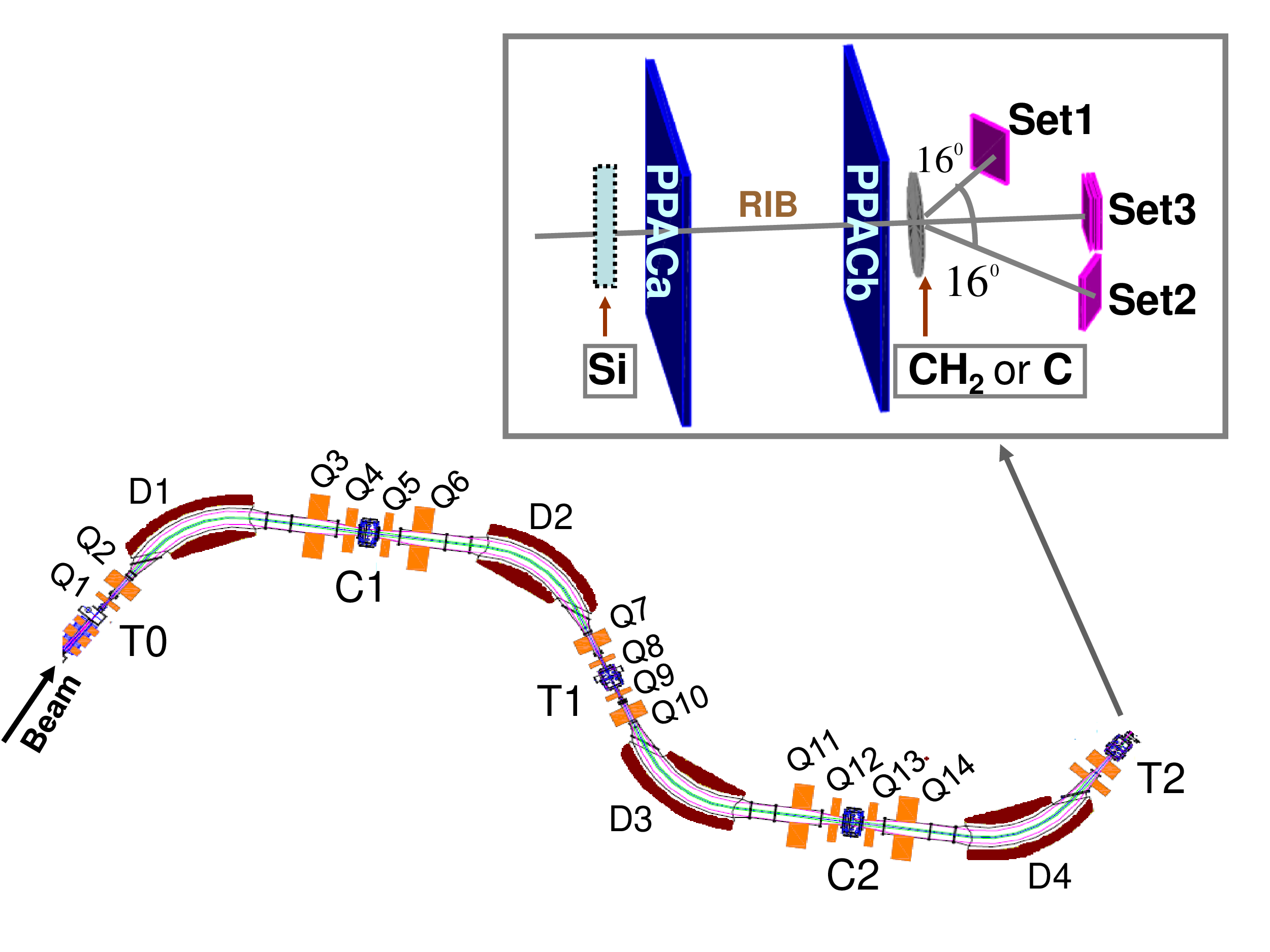}}
\vspace{-0.8cm}
\caption{(Color online) The schematic view of RIBLL and the experiment setup for the measurement of $^{10}$Be + p reaction. See text for details. }
\label{expset}
\vspace{+0.0cm}
\end{figure}

The measurement of $^{10}$Be + p elastic resonance scattering was carried out at the Heavy Ion Research
Facility in Lanzhou (HIRFL) \cite{Zhan2008, Zhan2010}, China. A secondary ion beam of $^{10}$Be was
produced by the Radioactive Ion Beam Line in Lanzhou (RIBLL) through the projectile fragmentation
of a 59.62$A$ MeV $^{18}$O primary beam with an average intensity of 100 enA bombarded on a solid  853.96 mg/cm$^{2}$-thick target of $^{9}$Be. The schematic view of RIBLL and the experiment setup for the present measurement are shown in Fig. \ref{expset}. At the first momentum-dispersive focal plane (C1), a 3060 $\mu$m-thick aluminium achromatic degrader was installed to reject unwanted ion species from the $^{10}$Be beam. Two nominal horizontal slits at focal planes C1 (S1: $\pm$ 15 mm) and C2 (S2: $\pm$ 25 mm) were used to restrict the momentum spread of $^{10}$Be beam. The time of flight (TOF) from the first achromatic focal point T1 to the second one T2 was measured by two 50 $\mu$m-thick plastic scintillators.

\begin{figure}[htbp]
\resizebox{10.6cm}{!}{\includegraphics{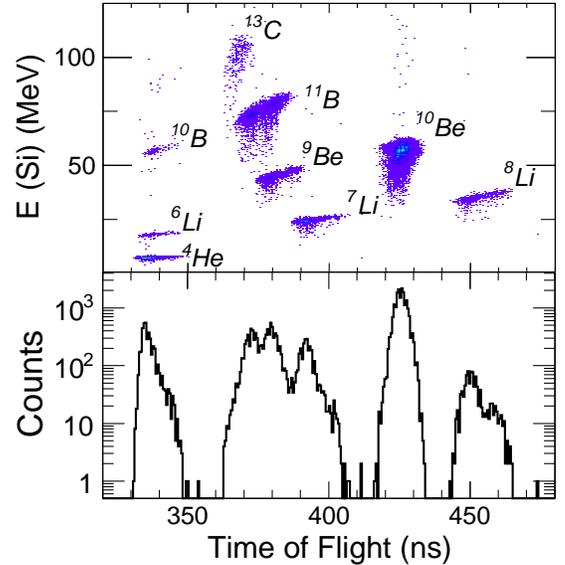}}
\centering
\vspace{-0.8cm}
\caption{(Color online) The particle identification plot of E(Si) versus TOF detectors (top panel) and its projection on the TOF-axis (bottom panel). E is the energy deposited in the Si detector with 280 $\mu$m thickness, and TOF is the time of flight (ns) of particles from T1 to T2.}
\label{de_tof}
\vspace{+0.0cm}
\end{figure}

At the scattering chamber (see Fig. \ref{expset}), the incident angles of beam particles on the target were determined by extrapolating the hit-positions recorded by two parallel-plate avalanche counters (PPAC), i.e., PPACa and PPACb, which were 500 mm and 100 mm away from the target, respectively. Consequently, the $^{10}$Be beam of about 92\% purity with an intensity of about 5$\times$10$^{3}$ pps and an energy of 7.2$A$ MeV impinged on a 41.85 mg/cm$^{2}$-thick polyethylene (PE) target. In addition, a 280 $\mu$m-thick silicon detector (Si) was inserted in front of PPACa during the beam tuning to perform the particle identification with TOF. As shown in Fig. \ref{de_tof}, $^{10}$Be nuclei can be distinguished unambiguously with the help of TOF.

\begin{figure}[htbp]
\resizebox{8.6cm}{!}{\includegraphics{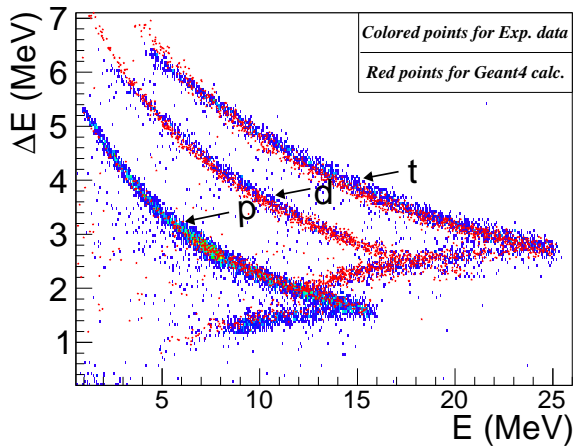}}
\vspace{-0.8cm}
\caption{(Color online) The scattering plot of recoiled light particles (colored points) from Set1 $\Delta$E-E telescope, with the locus of protons, deuterons, and tritons labeled by arrows. The simulation results (red points) of the present reaction by GEANT4~\cite{Agostinelli2003} are also shown for comparison.}
\label{partidwithg4}
\vspace{+0.0cm}
\end{figure}

The scattered particles from the reaction target were detected by three downstream $\Delta$E-E silicon
telescopes at the end of the scattering chamber, which were placed at laboratory angles of $\theta_{lab}$
$\approx$ +16$^{o}$, -16$^{o}$, and 0$^{o}$ with respect to the beam axis (see Fig. \ref{expset}).
The distances from the telescopes to the reaction target, with solid angles of 21 msr, 20 msr, and
21 msr, were set to be about 344 mm, 349 mm, and 345 mm, respectively. For the sake of convenience in the
following description, hereafter the three $\Delta$E-E silicon telescopes are referred to Set1, Set2, and Set3.
The averaged scattering angles for recoiled particles were determined to be
$\theta_{c.m.}$ $\approx$ 148$^{o}$, 147$^{o}$, and 180$^{o}$ via the relation of
$\theta_{c.m.}$ = 180$^o$ - 2$\theta_{lab}$. The $\Delta$E detectors were
position-sensitive double-sided-strip detectors (16 $\times$ 16, 3 mm width of each strip),
measuring  energy and two-dimensional position information of the recoiled particles with
thickness of 301 $\mu$m, 300 $\mu$m, and 149 $\mu$m, respectively. The E silicon detectors, with thickness of 1533 $\mu$m and 1524 $\mu$m in Set1 and Set2, and 1528 $\mu$m + 1538 $\mu$m in Set3,
measured the partial or whole energy, depending on whether the scattered particles punching through
the $\Delta$E-E telescopes. As demonstrated in Fig. \ref{partidwithg4}, the combination
of $\Delta$E-E detectors clearly revealed the ability of separating protons from deuterons and tritons while heavier ions were stopped in the target. The time of flight and Si detectors were calibrated using the primary beam $^{18}$O, while the position calibration for PPACs and the energy calibration for three silicon telescopes were performed utilizing a standard double $\alpha$ source of energy 5.156 MeV from $^{239}$Pu and 5.499 MeV from $^{238}$Pu. Since the two points of $\alpha$ source are very close in energy, the back bending points for protons and tritons in the $\Delta$E (channel)-E (channel) plot were utilized to carry out the calibration for three silicon telescopes. As shown in Fig. \ref{partidwithg4}, the calibration result was in excellent agreement with that from GEANT4 \cite{Agostinelli2003} simulations.

Additionally, in the last stage of the measurement, an evaluation of background contribution from the reactions of $^{10}$Be with carbon nuclei in the (CH$_{2}$)$_n$ target was performed through a separate run with a 50.69 mg/cm$^{2}$-thick carbon target and the same $^{10}$Be beam as described above. In the following spectra, the data was directly extracted from the locus of protons as shown in Fig. \ref{partidwithg4} except for the high-energy ones that were ceased by depositing the remaining energy into detector telescopes. The proton yield ratio of two runs was normalized by the number of beam particles and target thickness per unit energy loss of the incident beam in the target.

\section{results and discussions}
\label{sec:resultanddiss}

\begin{figure}[htbp]
\resizebox{8.6cm}{!}{\includegraphics{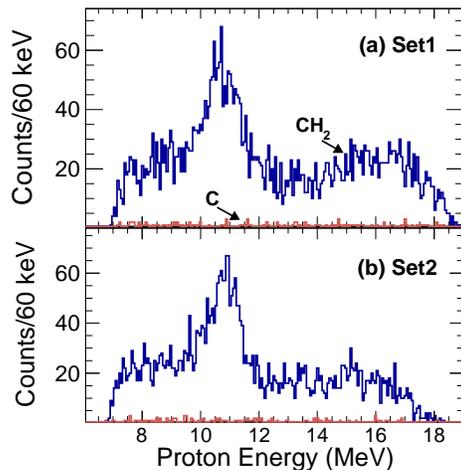}}
\vspace{-0.8cm}
\caption{(Color online) The energy spectra of protons for Set1 (a) and Set2 (b), in which the energy straggling in the target have been corrected. The contributions of protons from the reactions of $^{10}$Be with carbon nuclei in the (CH$_2$)$_n$ target are indicated by red curves.}
\label{Ep}
\vspace{+0.0cm}
\end{figure}

For the elastic resonance scattering, the relation between the energy of recoiled protons detected at a laboratory scattering angle $\theta_{sca}$ with the center-of-mass (C.M.) energy, is expressed as:
\begin{equation}
E_{c.m.}=\frac{A_P+A_T}{4A_P\cos^{2}\theta_{sca}}E_p,
\label{ecm}
\end{equation}
where A$_{P}$ and A$_{T}$ are the mass numbers of projectile and target nuclei, respectively, and E$_{p}$ is total proton
energy derived using an energy loss program \cite{Ziegler1980} within an uncertainty of about 5\%.
The energy loss of particles in the target was deduced from a Monte Carlo simulation. The $\theta_{sca}$ represents the scattering angle between the beam direction and the outgoing direction of protons. The proton energy spectra for Set1 and Set2 are shown in Fig.~\ref{Ep}. The deduced E$_{c.m.}$ resolution, consisting of the energy width of the secondary beam, the energy resolution of the silicon telescope, the angular resolution of the scattering angle, and the energy straggling in the target, is estimated to be 40 $\sim$ 100 keV, depending on the relative energies of the reaction system.

It is possible that the proton-decays from compound nuclei $^{11}$B to the first excited state
(3.368 MeV, 2$^{+}$) or higher excited states in $^{10}$Be contribute to proton spectra.
However, according to the available data of elastic and inelastic proton scatterings from a $^{10}$Be
target \cite{Auton1970} and using a distorted-wave Born approximation (DWBA)  to the energy of interest, the inelastic scattering
contributions in our case from the channels of p + $^{10}$Be$^{\ast}$ can be neglected, thus the energy
derived from Eq. (\ref{ecm}) can be reasonably recognized as the elastic scattering events.
In addition, the background contribution of protons from the reactions of $^{10}$Be with carbon nucleus
in the (CH$_2$)$_n$ target is found to be negligible in comparison with those of the p + $^{10}$Be reactions as shown in Fig. \ref{Ep}.

The achieved differential cross section through a transformation from the laboratory to the C.M. frame energy is presented by
\begin{equation}
\left(\frac{d\sigma}{d\Omega}\right)_{c.m.}(\theta_{c.m.}, E_{c.m.})=\frac{1}{4\cos\theta_{lab}}\left(\frac{d\sigma}{d\Omega}\right)_{lab}(\theta_{lab}, E_{p}),
\label{lab2cmdiffcro}
\end{equation}
where E$_{c.m.}$ is the C.M. energy of p + $^{10}$Be system calculated by Eq. (\ref{ecm}). The quantity on the right-hand side of Eq. (\ref{lab2cmdiffcro}) is expressed as
\begin{equation}
\left(\frac{d\sigma}{d\Omega}\right)_{lab}(\theta_{lab}, E_{p})=\frac{N_{p}}{N_{0}N_{t}\Delta\Omega},
\label{labdiffcro}
\end{equation}
representing the nominal laboratory differential cross section at recoiled proton energy E$_{p}$ and laboratory angle $\theta_{lab}$ for silicon telescopes placed with respect to the reaction target, with N$_{0}$ being the total number of $^{10}$Be bombarded on the (CH$_{2}$)$_{n}$ target, N$_{t}$ being the number of hydrogen atoms per unit area (cm$^{2}$) per energy bin (dE$_{p}$) in the target, and N$_{p}$ being the number of protons detected per energy bin by the $\Delta$E-E telescope with a solid angle $\Delta\Omega$.

\begin{figure}[htbp]
\centering
\resizebox{10.9cm}{!}{\includegraphics{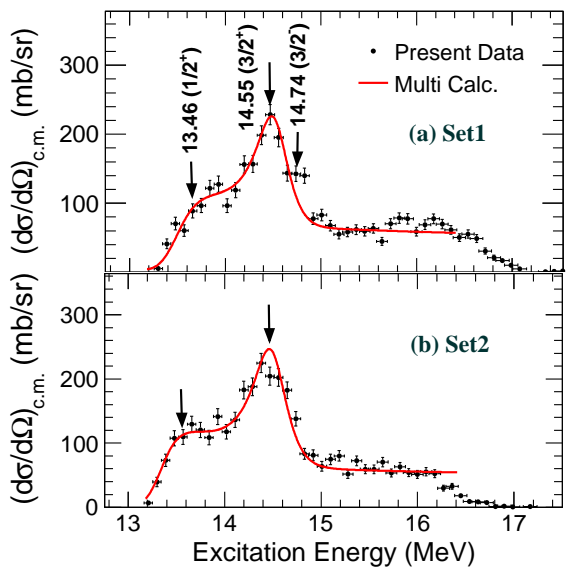}}
\vspace{-0.8cm}
\caption{(Color online) The extracted excitation function of cross sections for Set1 (a) and Set2 (b), with the vertical error bars
 showing statistical errors only, the horizontal error bars representing systematic uncertainty of energy levels, and arrows indicating the populated resonant states with energy level in MeV. The solid curves are the R-matrix fits to the data. See text for details.}
\label{excitfun}
\vspace{+0.2cm}
\end{figure}

The excitation energies in $^{11}$B are deduced using the relation of E$_{x}$ = E$_{r}$ + Q$_p$ where the proton separation energy Q$_p$ is 11.2285 MeV and the resonant energy E$_r$ is obtained by the following R-matrix analysis of the data, the resulting proton excitation functions are depicted in Fig. \ref{excitfun} in which the background contributions of protons from the contamination of other reactions were subtracted. Small discrepancy between the two data sets respectively shown in Fig. \ref{excitfun} (a) and (b) can be seen, and it mainly originates from the counting statistics, the finite size of the detectors, and the contamination of deuterons and tritons as shown in Fig. \ref{partidwithg4}.

\begin{table*}[htbp]
\caption{The extracted resonant level parameters from the R-matrix best-fit to the present data. Those tentative assignments of J$^\pi$ or probably populated states are put in parentheses. For comparison, the extracted total or partial widths of states ($\Gamma_{p}$) and corresponding widths of Wigner limit ($\Gamma_{W}$) are shown in unit of keV. Errors of overall uncertainty for energy levels and fitting uncertainty for resonant partial widths are given.}
\label{TBestPar}
\flushleft
  \begin{tabular}{p{100pt}   p{100pt}    p{60pt}   p{65pt}  p{90pt}   p{70pt} }
  \hline
  \hline
          E$_x$ (MeV)        &  E$_{r}$ (MeV)   & l$_p$  &  J$^\pi$            & $\Gamma_{p}$ (keV)  & $\Gamma_{W}$ (keV)\\
          \hline
          13.46 $\pm$ 0.13       &  2.23 $\pm$ 0.13    &  0     &  $1/2^+$            & 608 $\pm$ 242       &  1537 \\
          14.55 $\pm$ 0.07       &  3.32 $\pm$ 0.07    &  2     &  $3/2^+$            & 475 $\pm$ 80        &  547\\

          14.74 $\pm$ 0.09       &  3.51 $\pm$ 0.09    &  1     &  $3/2^-$            & 830 $\pm$ 145       &  1507\\
          (16.18) $\pm$ 0.21     &  4.95 $\pm$ 0.21    &  1     &  ($1/2^-$, $3/2^-$) & $<60$               &  206\\
  \hline
  \hline
  \end{tabular}
\end{table*}

Analysis of the data was performed by a multichannel R-matrix program $Multi$ \cite{Lane1958, Brune2002, Descouvemont2010} to extract resonant parameters such as the resonant energy E$_{r}$, the spin-parity J$^{\pi}$, and the proton-decay partial width $\Gamma_{p}$. In the present work, a channel radius of 4.606 fm given by R$_{c}$ = 1.46(A$_T$$^{1/3}$ + A$_P$$^{1/3}$) has been utilized. The fit results were insensitive to the choice of the radius in the present energy region of interest. The inelastic channel and $\gamma$ capture channel widths are neglected. Therefore, the single particle width is derived by $\Gamma_{sp}$ = 2P$_{l}\gamma^2$, where P$_l$ is the barrier penetrability factor and $\gamma^2$  is the reduced particle partial width. In practice, the calculated particle widths are frequently compared with the Wigner limit width as a measure of the partial width of a resonance in terms of $\Gamma_W$ = 2P$_{l}\gamma_W ^2$ with $\gamma_W^2$ = 3$\hbar^2$/2$\mu$R$_c^2$ in which $\mu$ is the reduced mass \cite{Teichmannand1952}.

Since the spin-parity of protons and the ground state of $^{10}$Be are 1/2$^+$ and 0$^+$, respectively, the incident channel spin is determined to be $s = 1/2$. Regarding conservation of the total angular momentum $J$ obtained with relative orbit angular momentum $l$ coupling to the channel spin $s$, many R-matrix fits with all possible spin-parity combinations for observed resonances were attempted to reproduce the data. Consequently, the best-fit parameters are summarized in Table \ref{TBestPar} and the resulting fits are displayed in Fig. \ref{excitfun}. In the following subsections, the detailed discussions for the observed states are presented.

\subsection{Bump around 13.46 MeV and Peak at 14.55 MeV}

\begin{figure}[htbp]
\resizebox{9.2cm}{!}{\includegraphics{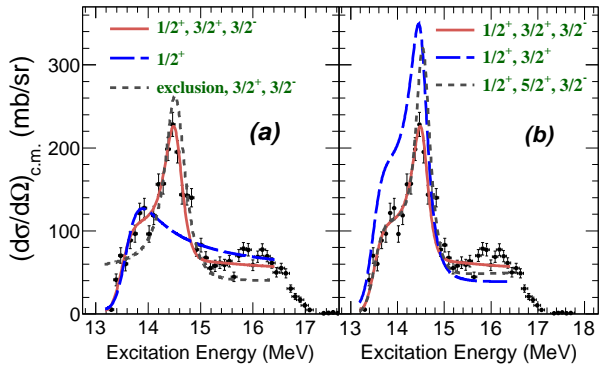}}
\vspace{-0.8cm}
\caption{(Color online) R-matrix fits to the present data with resonant energy levels near 13.46 MeV (left panel), 14.55 MeV, and 14.74 MeV (right panel).}
\label{levelsfit}
\vspace{+0.0cm}

\end{figure}

Because of the close proximity and the proton-decay channel, the peak at E$_x$ = 14.55 MeV is assumed to represent a resonance as the well-known state at 14.34 MeV, and the large proton-decay width of this resonance precludes the possibility of the previously known state at 14.56 MeV \cite{Kelly2012}. Attempts were made to fit the peak around E$_x$ = 14.50 MeV with J$^\pi$ = 1/2$^+$, 3/2$^\pm$, and 5/2$^+$, which are assignments made by previous measurements \cite{Goosman1973, Zwieglinski1982, Zwieglinski1979, Soic2004}. Calculations were also guided by the most recently tabulated levels for $^{11}$B \cite{Kelly2012} and the isobaric analog states in $^{11}$Be. This state was firstly observed at E$_x$ = 14.33 MeV in the $^{10}$Be(p, $\gamma$)$^{11}$B reaction \cite{Goosman1973} and identified as the analog state of 1.78 MeV in $^{11}$Be with the most probable J$^\pi$ = 5/2$^+$ assignment~\cite{Pullen1962}, but the possibility of J$^\pi$ = 3/2$^+$ was firmly excluded on the basis of quadrupole-to-dipole amplitude analysis. Later, other studies on this state were totally based on Goosman's measurement. Therefore, the preferred assignment for this state is adopted to be J$^\pi$ = 5/2$^+$ in the recent compilation of A = 11 nuclei \cite{Kelly2012}. Our result demonstrates that the curve with J$^\pi$ = 3/2$^+$ assignment is in accordance with the data shown in Fig. \ref{levelsfit}(b) as the solid curve, while the combination of J$^\pi$ = 5/2$^+$ with other states yields good description of the overall shape except for the height of the prominent peak shown in Fig. \ref{levelsfit}(b) as the black dotted curve.

Most of previous studies have measured the state J$^\pi$ = 5/2$^+$, and excitation energies of E$_x$ = 14.34 \cite{Goosman1973}, 14.47 \cite{Soic2004}, 14.40 \cite{Watson1971,Zwieglinski1982}, and 14.30 MeV \cite{Fletcher2003} have been reported. In the current measurement, the experimental resolution and counting statistics prevent us from making a firm assignment, thus the candidate J$^\pi$ = 5/2$^+$ can not be excluded. Instead, it may be a strong candidate as suggested by the DWBA and shell-model calculations in Refs. \cite{Zwieglinski1982, Teeters1977, Kohen1970}. A high-resolution study is required to make a definite assignment for this state.

As shown in Figs. \ref{excitfun} and \ref{levelsfit}, there is a bump in the low-energy region of
the excitation function. This structure was the result of low-energy cutoff
imposed by the $\Delta$E-E telescope except for the first two points, since no protons with energy lower
than E$_x$ = 13.0 MeV could be detected. As shown in Fig. \ref{levelsfit}(b), exclusion of this bump around 13.46 MeV results in a much less satisfactory fit by the blue dashed line for the overall shape of the spectrum. To improve the fitting, the bump is introduced as a resonance. Attempts were made to fit the bump with $J^\pi$ = $1/2^{\pm}$, $3/2^{\pm}$, and 5/2$^{\pm}$. As demonstrated by the blue dashed curve in Fig. \ref{levelsfit}(a), the bump can be well described by a Breit-Wigner form \cite{Freer2012M} with transferred angular momentum l$_p$ = 0. The excitation functions were well reproduced by using an assignment of J$^\pi$ = 1/2$^{+}$ for this state in combination with other states observed, but the extracted proton-decay width was roughly estimated to be $\Gamma_{p}$ = 608 keV with large uncertainty due to the paucity of the low-energy data.

For excitation energies at E$_x$ = 13.30 MeV \cite{Groce1963} and 13.63 MeV \cite{Cusson1966}, no information of the spin-parity and the particle width has been reported so far. In this work, the resonance at E$_x$ = 13.46 MeV was tentatively assumed to be the 13.30 MeV or 13.63 MeV state but not the member of the doublet, i.e., the 13.137 MeV (9/2$^+$) state and 13.16 MeV (5/2$^+$, 7/2$^{+}$) state \cite{Zwieglinski1982}, due to the restriction of the largest transferred angular momentum (l$_{max}$$\leq3$). Alternatively, this resonant state may be an unnatural-parity state that was not observed previously. Further studies above the $^{10}$Be + p threshold are needed to obtain a firm assignment for this state.

\begin{table*}[htbp]
\caption{The energy levels of $^{11}$B with proton-decay channels observed by the previous and present reactions in the energy range covered by the present measurement.}
\label{B11table}
\flushleft
          \begin{tabular}{p{130pt}   p{90pt}    p{115pt}      p{156pt} }
          \hline
          \hline
       E$_x$ (MeV)  &  J$^\pi$            &  $\Gamma_{p}$ (keV)   &Reactions (year)\\
          \hline
          13.30     &                     &                 &  (1963) \cite{Groce1963}\\
          13.63     &                     &                 & $^{7}$Li($\alpha$, $\alpha$$^\prime$)$^{7}$Li, $^{7}$Li($\alpha$, $\alpha$)$^{7}$Li (1966) \cite{Cusson1966}\\
          $\textbf{13.46 $\pm$ 0.13}$     &1/2$^+$          & 608 $\pm$ 242 & \textbf{this work}\\

          \hline
          14.34$^*$ &   5/2$^+$           &  253            &   \cite{Kelly2012}\\
          14.33     &  (5/2$^+$, 3/2$^-$) &  255            &   $^{10}$Be(p, $\gamma$)$^{11}$B (1973) \cite{Goosman1973}\\
          14.33     &  (1/2$^+$, 3/2$^+$, 5/2$^+$)$^a$&     &   $^9$Be(t, p)$^{11}$Be (1962) \cite{Pullen1962}\\
          1.78$^b$  &  (3/2$^+$, 5/2$^+$) &                 &   $^{10}$Be(d, p)$^{11}$Be (1979) \cite{Zwieglinski1979}\\
          14.40     &  (5/2$^+$)$^c$      &  261            &   $^{9}$Be($^3$He, p), $^{9}$Be($\alpha$, d) (1982) \cite{Zwieglinski1982}\\
          14.47     &                     &                 &   $^{16}$O($^9$Be, $\alpha^7$Be), $^7$Li($^9$Be, $\alpha$ $^7$Li) (2004) \cite{Soic2004}\\
          14.40     &                     &  220 $\pm$ 50.   &   $^{11}$B(t, t$^{\prime}$)$^{11}$B (1974) \cite{Watson1971}\\
          14.30     &                     &                 &   $^{7}$Li($^7$Li, t)$^{11}$B (2003) \cite{Fletcher2003}\\
          $\textbf{14.55 $\pm$ 0.07}$   &   (3/2$^+$, 5/2$^+$)  & 475 $\pm$ 80 & \textbf{this work}\\

          \hline
          15.29$^*$ &  (3/2$^-$)          &  282            &   \cite{Kelly2012}\\
          15.3      &                     & 635 $\pm$ 180  &   $^{10}$Be(p, $\gamma$)$^{11}$B (1973) \cite{Goosman1973}\\
          15.12     &                     &  750            &   $^{10}$B(n, n$^{\prime}$)$^{10}$B, $^{7}$Li($\alpha$, n)$^{10}$B (1968) \cite{Ajzenberg1968}\\
          15.2      &  (3/2$^+$, 5/2$^+$, 7/2$^+$)$^d$      &   700  &$^{10}$B(n, n$^{\prime}$)$^{10}$B (1974) \cite{Hausladen1973}\\
          15.29     &  (3/2$^-$, 5/2$^-$) &   282           &   $^{14}$C(p, $\alpha$)$^{11}$B (1985) \cite{Aryaeinejad1985}\\
          $\textbf{14.74 $\pm$ 0.09}$     &   3/2$^-$       &   830 $\pm$ 145    & \textbf{this work} \\

          \hline
          16.43$^*$ &  (5/2$^-$)          &   $<$ 30        &   \cite{Kelly2012}  \\
          16.43     &                     &                 &   $^{9}$Be(d, p)$^{10}$Be (1974) \cite{Annegarn1974}\\
          16.5, 16.2&                     &                 &   $^{11}$B($\gamma$, p)$^{10}$Be (1970) \cite{Sorokin1970}\\
          16.5      &                     &                 &   $^{11}$B($\gamma$, n)$^{10}$B (1965) \cite{Hayward1965}\\
          16.44     &                     &   $\leq$ 30     &   $^{9}$Be($^3$He, p), $^{9}$Be($\alpha$, d) (1982) \cite{Zwieglinski1982}\\

          16.5$^e$  &  (5/2$^-$)          &   201           &   $^{14}$C(p, $\alpha$)$^{11}$B (1985) \cite{Aryaeinejad1985}\\

          \textbf{(16.18 $\pm$ 0.21)}     &   (1/2$^-$, 3/2$^-$) &   $<$ 60   & \textbf{this work}\\

          \hline
          \hline
       \end{tabular}
 $^{*}$ Energy levels from the compilation for A = 11 light nuclei \cite{Kelly2012}.

 $^{a}$ Taken from the $^9$Be(t, p)$^{11}$Be experiment at 10 MeV and 14 MeV triton energy, where the observed E$_x$ = 1.78 MeV state in $^{11}$Be corresponds to the analog state in $^{11}$B at E$_x$ = 14.33 MeV with J$^\pi$ = (3/2$^-$, 5/2$^+$) \cite{Pullen1962}.

 $^{b}$ The tentative assignment of J$^\pi$ = (5/2$^+)$ for the state at E$_x$ = 1.78 MeV in $^{11}$Be (corresponding to the analog state E$_x$ = 14.33 MeV in $^{11}$B) from the comparison of the $^{10}$Be(d, p)$^{11}$Be experimental data with the DWBA calculation \cite{Zwieglinski1979} can not rule out the possibility of 3/2$^+$ state since the purity of  T = 3/2 in $^{11}$B (14.33 MeV state) is in debate.

 $^{c}$ This value is adopted from the $^{10}$Be(p, $\gamma$)$^{11}$B reaction \cite{Goosman1973}.

 $^{d}$ The J$^\pi$ = (1/2$^+$, 11/2$^+$) were excluded since they give too low and too high peak heights over the energy region \cite{Hausladen1973}.

 $^e$ It may not be the 16.44 MeV state \cite{Zwieglinski1982}. See also Ref. \cite{Kurath1975}.
 \end{table*}

\subsection{Bump around 14.74 MeV}

As shown in Fig. \ref{excitfun}(a), there is a bump around 14.74 MeV. The fitting to the data with exclusion
of the bump as a resonance results in an obvious deviation as shown in Fig. \ref{levelsfit}(b) by the blue dashed curve. In previous experiments, the characteristics of this resonance were observed at 15.30 MeV \cite{Goosman1973}, 15.29 MeV \cite{Aryaeinejad1985}, and 15.2 MeV \cite{Hausladen1973}, and were considered as the analog state of 2.78 MeV in $^{11}$Be. Earlier work associated with this state at 15.12 MeV was the studies of $^{10}$B(n, n$^{\prime}$)$^{10}$B, $^7$Li($\alpha$, n)$^{10}$B, and $^{10}$B(n, $\alpha$)$^7$Li reactions compiled in Ref. \cite{Ajzenberg1968}. However, no more information on spin-parities among these works was reported except for Refs. \cite{Hausladen1973} and \cite{Aryaeinejad1985}.

Based on the analysis of R-matrix fits to the data of neutron scattering from $^{10}B$ \cite{Hausladen1973},
Hausladen {\it et al.} suggested an assignment of J$^\pi$ = (3/2$^+$, 5/2$^+$, 7/2$^+$), where fits with
a p-wave state did not reproduce any distributions of the Legendre polynomial expansion coefficients.
Since then, Aryaeinejad {\it et al.} suggested a spin-parity assignment J$^\pi$ = $3/2^-$ through the study of $^{14}$C(p, $\alpha$)$^{11}$B
reaction \cite{Aryaeinejad1985} in which only negative-parity states with T = 3/2 were strongly excited
in $^{11}$B. In the current measurement, we speculated that this bump as a resonance was populated,
  and attempts with various combinations of excited energies and spin-parities were made to derive the resonant parameters. Consequently, the data was found to be reproduced in a satisfactory way with the J$^\pi$ = 3/2$^-$ assignment. Obviously, our result supports Aryaeinejad's assignment but with a larger particle width because of the present systematic resolution.

\subsection{Alternative Structure around 16.18 MeV}

In the previous measurements as listed in Table \ref{B11table}, the state with E$_{x}=16.432$ MeV and J$^\pi$ = (5/2$^-$) of $^{11}$B compiled in Ref. \cite{Kelly2012} was reported and has the resonant energy of 16.44 MeV in Ref.~\cite{Zwieglinski1982} and 16.43 MeV in Refs.~\cite{Annegarn1974, Sorokin1970}, respectively. R-matrix fits to the data were attempted for this state formed by l$_p$ = 0, 1, 2, and 3 protons. Calculations with p-wave states using a particle width smaller than 60 keV could naively produce a character of the tail of the excitation function, but the $\chi^2$ of the fit to the data was almost unchanged with the exclusion of this state. It is assumed that we have populated the well-known resonance, i.e., the analog of the 3.889 MeV state in $^{11}$Be, and a tentative assignment of J$^\pi$ = (1/2$^-$, 3/2$^-$) is set to this state. The characteristic behavior of angular momentum transfer l$_p$ = 1 for this state disagrees with the previously observed one at 16.50 MeV \cite{Aryaeinejad1985} with a width of 201 keV, which was not represented by any analog states in $^{11}$Be.

\section{summary}
\label{sec:sum}

To summarize, the elastic resonance scattering of protons from $^{10}$Be using the thick-target technique in inverse kinematics was performed for the first time to investigate the resonant structure of $^{11}$B above the proton-decay threshold. The excitation function for the resonant protons decayed from the compound nucleus $^{11}$B was analysed using the multichannel R-matrix procedure. Four resonant states at E$_x$ = 13.46 MeV, 14.55 MeV, 14.74 MeV, and 16.18 MeV were populated, where the latter three levels are considered to be the previously known ones with isospin value T = 3/2, and the data was reproduced well with the combination of those resonant parameters as listed in Table.~\ref{TBestPar}. The extracted parameters seem reliable within uncertainty except for the 14.55 MeV resonance as implied by the large ratio of $\Gamma_p/\Gamma_W$ arising from the counting statistics. Alternatively, the 16.18 MeV state which is located at the end of the higher excitation energy is tentatively assigned to be the one observed previously. The extracted parameters for the resonant energy, the spin parity, and the proton-decay partial width are summarized in Table \ref{B11table}. The present results suffer from the large uncertainty in the level width, the spin-parity assignments, and even the isospin components, and further studies based on high-resolution experiments are imperative to draw firm conclusions for the structure of $^{11}$B.

$Acknowledgements:$ The authors would like to thank the HIRFL staffs for their operation of the accelerators and Jun Xu for helping with polishing the manuscript. This work was partially supported by the National Natural Science Foundation of China under Grant Nos. 11421505, 11075195, 11475245, and 11305239 and the Major State Basic Research Development Program in China under Contract No. 2013CB834405.

\end{document}